\documentclass{aa}    

\usepackage{graphicx,url,twoopt,natbib}
\usepackage[varg]{txfonts}           

\usepackage{pdfcomment}              
\usepackage{acronym}                 

\usepackage{url}
\usepackage{color,hyperref}
\definecolor{linkcolor}{rgb}{0,0.3,0.7}
\hypersetup{colorlinks=true,
	linkcolor=linkcolor, 
	citecolor=linkcolor,
	filecolor=linkcolor, 
	urlcolor=linkcolor}

\usepackage{natbib}

\bibpunct{(}{)}{,}{a}{}{,}    

\newcommand{\farc}{\hbox{$.\!\!^{\prime\prime}$}}

\newcommand{\lya}{Ly$\alpha$}

\newcommand{\hb}{H$\beta$} 
\newcommand{\ha}{H$\alpha$}

\newcommand{\oii}{[\ion{O}{ii}]} 
\newcommand{\oiii}{[\ion{O}{iii}]}

\usepackage{xcolor}

\begin{document}
	
\title{The host galaxy of the short GRB~111117A at $z = 2.211$: impact on the short GRB redshift distribution and progenitor channels\thanks{Based on observations collected at ESO/VLT under programme 088.A-0051 and 091.D-0904, at TNG under programme A24TAC\_38, at Gemini North under programme GN-2011B-Q-10 and GTC under programme GTC43-11B.}}

\titlerunning{GRB~111117A}

\author{J.~Selsing\inst{1}
\and T.~Kr\"{u}hler\inst{2}
\and D.~Malesani\inst{1}
\and P.~D'Avanzo\inst{3}
\and S.~Schulze \inst{4, 5, 6}
\and S.~D.~Vergani\inst{3, 7}
\and J.~Palmerio\inst{8}
\and J.~Japelj\inst{9}
\and B.~Milvang-Jensen\inst{1}
\and D.~Watson\inst{1}
\and P.~Jakobsson\inst{10}
\and J.~Bolmer \inst{2}
\and Z.~Cano\inst{11}
\and S.~Covino \inst{12}
\and V.~D'Elia\inst{12, 13}
\and A.~de~Ugarte~Postigo\inst{11, 1}
\and J.~P.~U.~Fynbo\inst{1}
\and A.~Gomboc\inst{14}
\and K.~E.~Heintz\inst{10,1}
\and L.~Kaper \inst{9}
\and A.~J.~Levan \inst{15}
\and S.~Piranomonte \inst{16}
\and G.~Pugliese \inst{9}
\and R.~S\'{a}nchez-Ram\'{\i}rez \inst{11, 17}
\and M.~Sparre\inst{1,18}
\and N.~R.~Tanvir\inst{19}
\and C.~C.~Th\"{o}ne\inst{11}
\and K.~Wiersema \inst{19}
}

\institute{Dark Cosmology Centre, Niels Bohr Institute, University of Copenhagen, Juliane Maries Vej 30, 2100 K\o benhavn \O, Denmark
\and Max-Planck-Institut f\"{u}r extraterrestrische Physik, Giessenbachstra\ss e, 85748 Garching, Germany
\and INAF - Osservatorio Astronomico di Brera, via E. Bianchi 46, I-23807, Merate (LC), Italy
\and Instituto de Astrof\'isica, Facultad de F\'isica, Pontificia Universidad Cat\'olica de Chile, Vicu\~{n}a Mackenna 4860, 7820436 Macul, Santiago, Chile
\and Millennium Institute of Astrophysics, Vicu\~{n}a Mackenna 4860, 7820436 Macul, Santiago, Chile
\and Department of Particle Physics and Astrophysics, Faculty of Physics, Weizmann Institute of Science, Rehovot 76100, Israel
\and GEPI, Observatoire de Paris, PSL Research University, CNRS, Place Jules Janssen, 92190 Meudon, France
\and Sorbonne Universités, UPMC Univ. Paris 6 et CNRS, UMR 7095, Institut d’Astrophysique de Paris, 98 bis bd Arago, 75014 Paris, France
\and Anton Pannekoek Institute for Astronomy, University of Amsterdam, Science Park 904, 1098 XH Amsterdam, The Netherlands
\and Centre for Astrophysics and Cosmology, Science Institute, University of Iceland, Dunhagi 5, 107 Reykjav\'ik, Iceland
\and Instituto de Astrof\'isica de Andaluc\'ia (IAA-CSIC), Glorieta de la Astronom\'ia s/n, E-18008, Granada, Spain
\and INAF-Osservatorio Astronomico di Roma, Via Frascati 33, I-00040 Monteporzio Catone, Italy
\and ASI-Science Data Centre, Via del Politecnico snc, I-00133 Rome, Italy
\and Centre for Astrophysics and Cosmology, University of Nova Gorica, Vipavska 13, 5000 Nova Gorica, Slovenia.
\and Department of Physics, University of Warwick, Coventry CV4 7AL, UK
\and INAF, Osservatorio Astronomico di Brera, Via E. Bianchi 46, I-23807 Merate (LC), Italy
\and Istituto de Astrofisica e Planetologia Spaziali, INAF, Via Fosso del Cavaliere 100, I-00133 Roma, Italy
\and Heidelberger Institut f{\"u}r Theoretische Studien, Schloss-Wolfsbrunnenweg 35, 69118 Heidelberg, Germany 
\and Department of Physics \& Astronomy and Leicester Institute of Space \& Earth Observation,
University of Leicester, University Road, Leicester, LE1 7RH, UK
}

\date{Received/ accepted}

\authorrunning{Selsing et al.}

\abstract{
It is notoriously difficult to localize short $\gamma$-ray bursts (sGRBs) and
their hosts to measure their redshifts. These measurements, however, are
critical to constrain the nature of sGRB progenitors, their redshift
distribution and the $r$-process element enrichment history of the universe.
Here, we present spectroscopy of the host galaxy of GRB~111117A and measure its
redshift to be $z = 2.211$. This makes GRB~111117A the most distant
high-confidence short duration GRB detected to date. Our spectroscopic redshift
supersedes a lower, previously estimated photometric redshift value for this
burst.

We use the spectroscopic redshift, as well as new imaging data to constrain the
nature of the host galaxy and the physical parameters of the GRB. The rest-frame
X-ray derived hydrogen column density, for example, is the highest compared to a
complete sample of sGRBs and seems to follow the evolution with redshift as
traced by the hosts of long GRBs (lGRBs).

The host lies in the brighter end of the expected sGRB host brightness distribution
at $z = 2.211$, and is actively forming stars. Using the host as a benchmark for
redshift determination, we find that between 43 and 71 per cent of all sGRB redshifts
should be missed due to host faintness for hosts at $z\sim2$. The high redshift of
GRB~111117A is evidence against a lognormal delay-time model for sGRBs through
the predicted redshift distribution of sGRBs, which is very sensitive to
high-$z$ sGRBs.

From the age of the universe at the time of GRB explosion, an initial neutron
star (NS) separation of $a_0 < 3.2~R_\odot$ is required in the case where the
progenitor system is a circular pair of inspiralling NSs. This constraint
excludes some of the longest sGRB formation channels for this burst.
}

\keywords{gamma-ray burst: individual: GRB~111117A -- gamma-ray burst: general -- galaxies: high-redshift -- binaries: general -- X-rays: bursts -- techniques: imaging spectroscopy}

\maketitle

\section{Introduction}

There is mounting evidence that short-duration $\gamma$-ray bursts
come from the merger of NSs, either with another NS, or a black
hole, due to their apparent association with kilonovae \citep{Barnes2013a,
	Tanvir2013b, Yang2015, Jin2016, Rosswog2016}. The absence of associated
supernovae in deep searches \citep[e.g.][]{Hjorth2005a,Fox2005,Hjorth2005b, Kann2011}
supports this idea and distinguishes the physical origin of sGRBs from their
long-duration counterparts, \citep[albeit see also][]{Fynbo2006b, Valle2006, Gal-Yam2006}.

The classification of GRBs in two groups, initially comes from the bimodal
distribution of burst duration and spectral hardness \citep{Kouveliotou1993},
where the duration T$_{90} < 2$ has been regarded as the dividing line between long
and short GRBs. Additionally, it has been found for lGRBs that there is a
spectral lag in the arrival-time of photons, with the most energetic ones
arriving first. This lag is consistent with zero for sGRBs
\citep{Norris2006}. Because both populations have continuous, overlapping
distributions in their observables and because telescopes observe in differing
bands, it is difficult to impose a single demarcation criterion between the two
classes. For this reason, the distinction between long and short GRBs is
preferably based on a combination of high-energy properties \citep{Zhang2009,
	Bromberg2012a, Bromberg2013}.

The \textit{Swift} satellite \citep{Gehrels2004} greatly improved the
understanding of sGRB progenitors thanks to its quick localization capability.
The bulk of these localizations have associated galaxies at relatively low
redshifts with a median redshift of $z\sim0.5$ \citep{Berger2014}, and because
most of these measurements come from the associated hosts, it is arguably biased
towards lower redshifts. The host galaxies of sGRBs are diverse. They are more
massive and less actively star-forming on average than lGRB hosts
\citep{Fong2013b}, while in some cases, no host galaxy can be identified above
detection threshold of deep follow-up observations \citep{Berger2010a,
	Tunnicliffe2014}. Together with their position within their hosts
\citep{Fong2013a}, this suggests a progenitor system that can be very long lived
in comparison to lGRBs, and is tracing with host stellar mass rather than
star-formation rate (SFR).

The electromagnetic signals from sGRBs are a promising channel to accurately
localize NS mergers, which holds the promise for a detection of an associated
gravitational wave (GW) signal \citep{Ghirlanda2016}. The simultaneous detection
of a sGRB and a GW will provide new promising ways to constrain the binary
inclination angle \citep{Arun2014} and measure cosmological distances
\citep{Nissanke2010}.

The total lifetimes of NS binaries depends on their orbit, mass, spin, initial
separations and subsequent inspiral times. The delay time from formation to
explosion impacts the timing and distribution of the enrichment of the ISM with
heavy $r$-process elements \citep{VandeVoort2015, Wallner2015,  Ji2016}. Some
limits can be calculated using host galaxy star-formation history models and
spatial distribution of sGRBs within their hosts \citep[][]{Berger2014}. The
most distant cosmological bursts, however, offer direct, hard limits on the
coalescence time scales.

We here present a spectrum of the host galaxy of the short
GRB~111117A ($T_{90}=0.46$~s) and measure its redshift to be $z=2.211$. This
value is significantly higher than the previously estimated redshift based on
photometric studies \citep{Margutti2012,Sakamoto2013}. We present the GRB's rest
frame properties based on this new distance compared to previous analyses and
revisit the host properties derived from the new solution to the spectral energy distribution (SED) fit.
While no optical afterglow was detected, the excellent localization from a
detection of the X-ray afterglow by the \emph{Chandra X-ray Observatory}
allows us to discuss the positioning and environmental properties of this
remarkably distant sGRB.
We use the $\Lambda$CDM cosmology parameters provided by
\citet{Planck2015} in which the universe is flat with $H_0 = 67.7$\,km\,s$^{-1}$
and $\Omega_m = 0.307$. All magnitudes are given in the AB system.


\section{Observations and results}

\subsection{Spectroscopic observations and analysis}

\begin{figure}
	\centering
	\includegraphics[width=9cm]{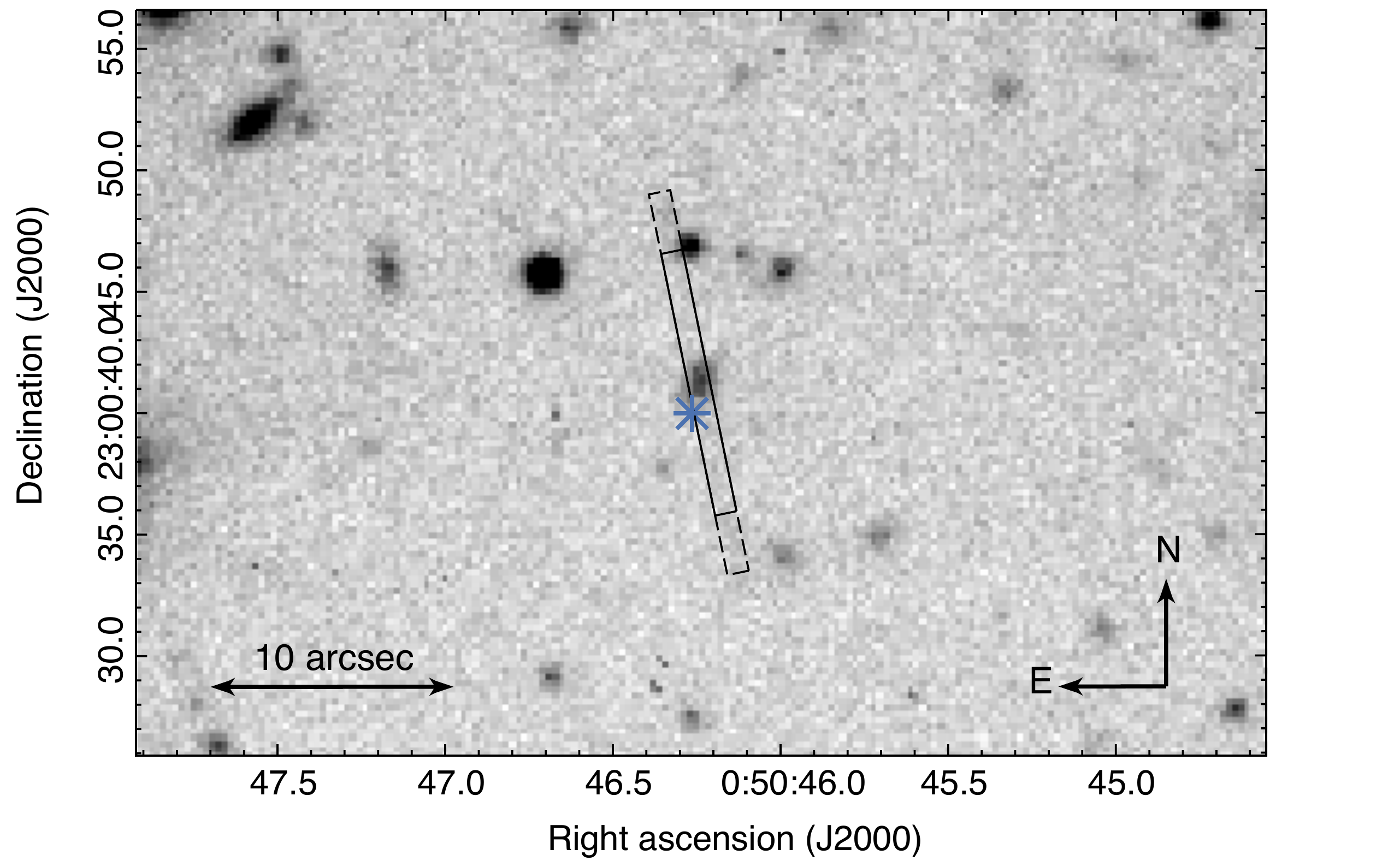}
	\caption{
	FORS2 R-band imaging of the field of GRB~111117A with the X-shooter slit overlaid. The slit
	position represents 4 epochs of spectroscopic observations taken at similar
	position angles. The corresponding photometry
	is shown in Fig.~\ref{fig:SED}. The blue asterisk indicates the GRB position as
	derived from the \emph{Chandra} observations in \citet{Sakamoto2013}. 
	}
	\label{fig:spec_setup}
\end{figure}

\begin{figure}
	\centering
	\includegraphics[width=9cm]{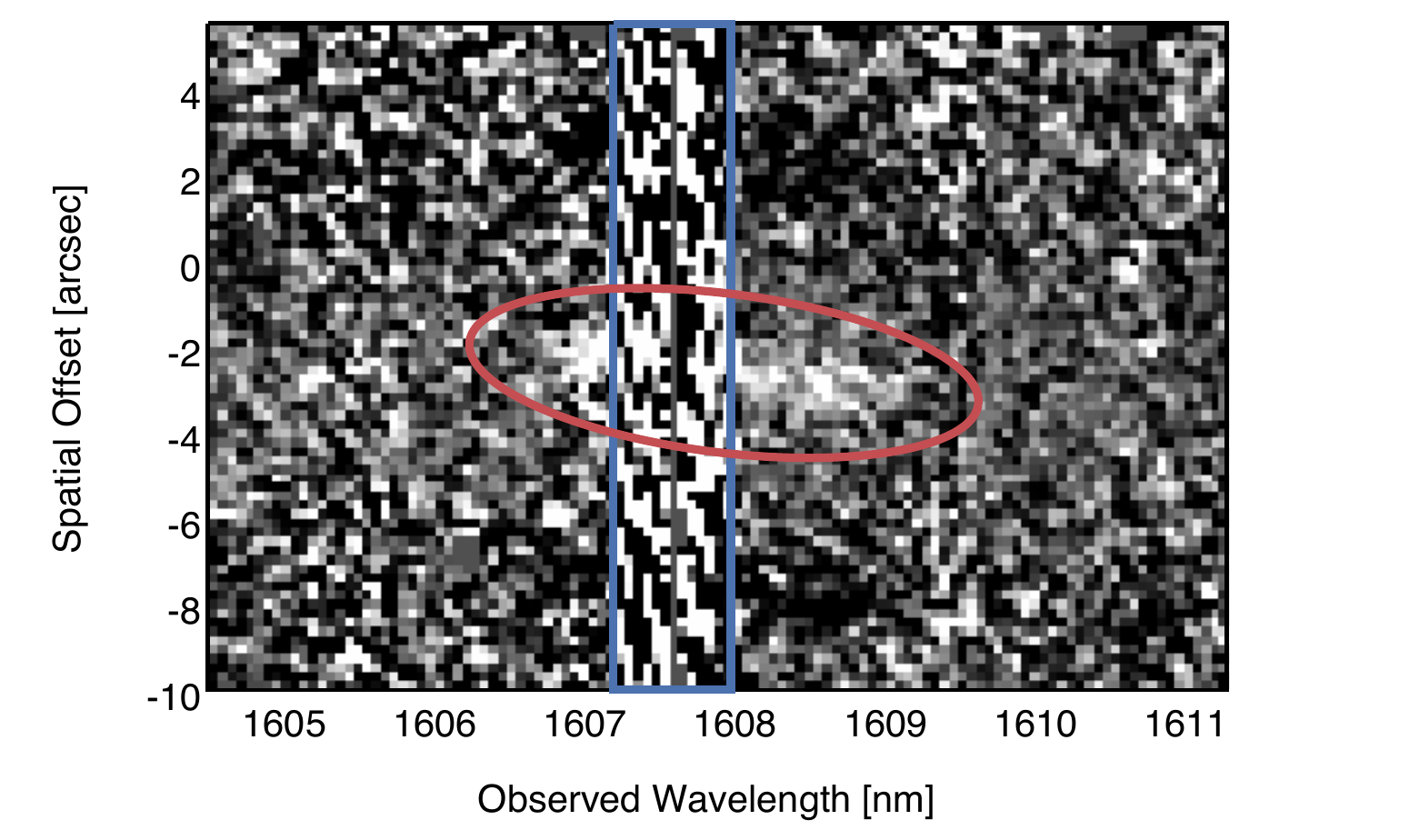}
	\caption{2D-image of the \oiii$\lambda$5007 emission line. The location of a bright skyline is marked by the blue box. The location of the emission line is indicated with the red ellipse. Because the host is observed in nodding-mode, negative images of the emission line appear on both sides in the spatial direction.}
	\label{fig:line}
\end{figure}

\begin{figure*}
	\centering
	\includegraphics[width=16cm]{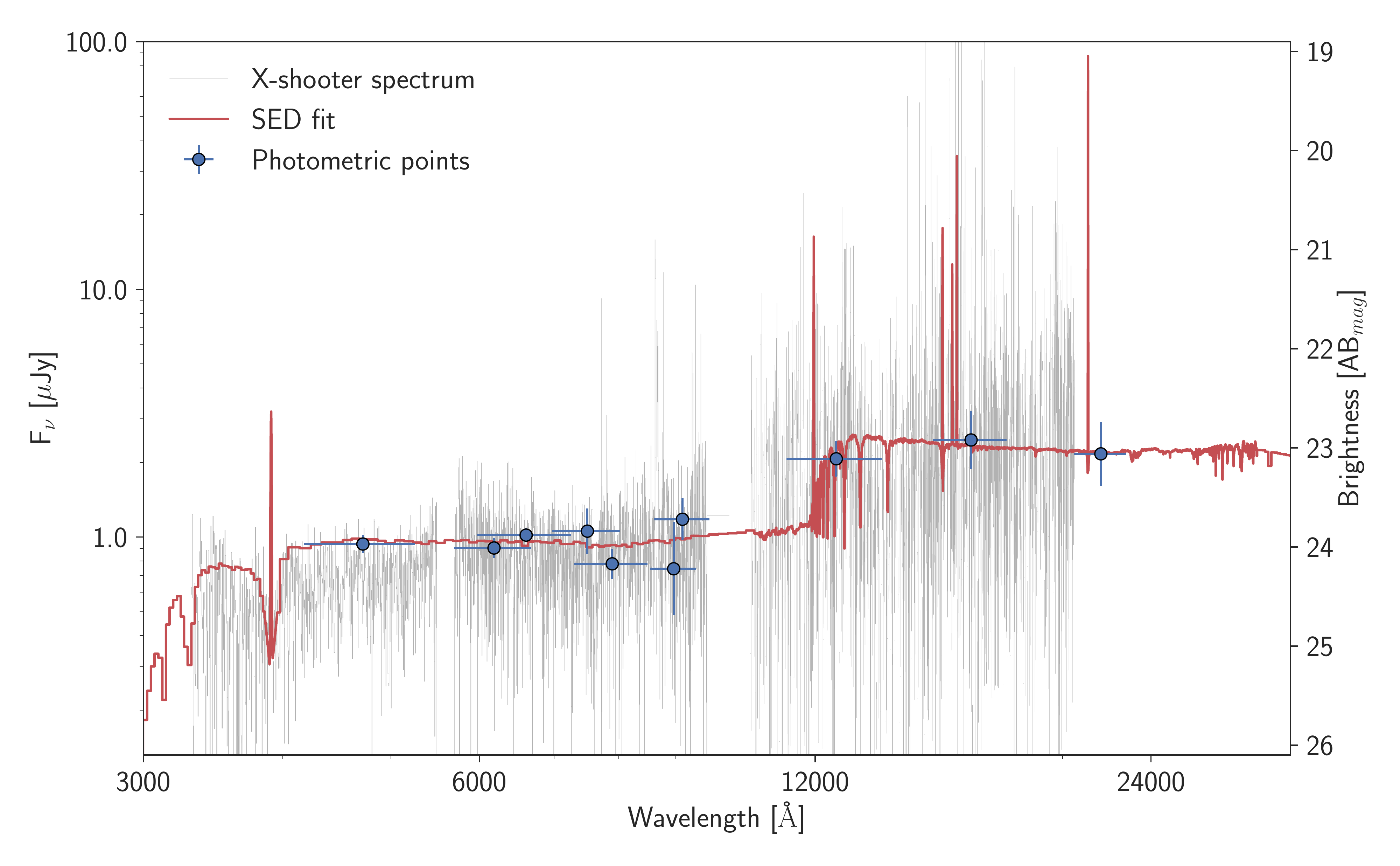}
	\caption{SED fit showning the best-fist SED to the derived photometry. The detection of \lya~ is predicted from the SED fit and confirmed by the spectroscopic observations. Overplotted in grey is the observed spectrum, binned down to 3 \AA~per pixel for presentation purposes. Slit losses are responsible for discrepancies between the measured photometry and the X-shooter spectrum. The reason for the spectral gaps at 5500 \AA~and 10000 \AA~is from the merging of the arms.}
	\label{fig:SED}
\end{figure*}

Spectroscopic observations were carried out using the cross-dispersed echelle
spectrograph, VLT/X-shooter \citep{Vernet2011}, at four seperate epochs. The
burst was observed 38 hours after the Burst Alert Telescope (BAT) trigger under
ESO programme 088.A-0051 (PI: Fynbo) and again later under ESO programme
091.D-0904 (PI: Hjorth). Observations use a simple ABBA nodding pattern, using 5
arcsec nod throws. X-shooter covers the wavelength range from 3000~\AA~to
24\,800~\AA~(21\,000~\AA~when the $K$-band blocking filter is used) across three
spectroscopic arms. We carried out the bias-correction, flat-fielding, order
tracing, wavelength calibration, rectification, and flux calibration using the
VLT/X-shooter pipeline version 2.8.4 \citep{Goldoni2006, Modigliani2010} run in physical
mode. Because the echelle orders are curved across each detector, a
rectification algorithm is employed which introduces correlations between
neighboring pixels. We select a pixel-scale of 0.2/0.2/0.6 \AA/pix for the
UVB/VIS/NIR arm to minimize the degree of correlation while conserving the
maximal resolution. The observations are combined and extracted using scripts
described in Selsing et al. 2017 (in prep.) and available
online\footnote{\url{https://github.com/jselsing/XSGRB_reduction_scripts}},
where the full spectral point spread function is modeled across each arm and
used for the optimal extraction algorithm \citep{Horne1986}. An overview of the
spectroscopic observations is given in Table~\ref{tab:spec_overview}, and the
slit position is shown in Fig.~\ref{fig:spec_setup}. We show the extracted
spectrum in Fig.~\ref{fig:SED}.

\begin{table*}[!ht]

	\centering
	\caption{Overview of the spectroscopic observations. JH in the slit width refers to observations where a K-band blocking filter has been used. The seeing is determined from the width of the spectral trace of a telluric standard star, taken close in time to the host observation. The spectral resolution, R, is measured from unresolved telluric absorption lines in the spectrum of the telluric standard star. \label{tab:spec_overview}}
	\begin{tabular}{cccccccc}
		\hline\hline
		{Obs. Date} &  \multicolumn{3}{c}{Exposure time (s)} & Slit width & Airmass & Seeing & R   \\ [1.5pt]
        \hline
		{} & UVB  & VIS & NIR &  (arcsec)   & {} & (arcsec)  & {VIS/NIR}  \\ [1.5pt]
		\hline
		2011-11-19T01:33 & 2 $\times$ 2400 & 2 $\times$ 2400 & 8 $\times$ 600 & 1.0/1.0/0.9 & 1.49 & 0.75 & 11600/6700 \\
        2013-07-15T09:02 & 2 $\times$ 1200 & 2 $\times$ 1200 & 8 $\times$ 300 & 1.0/1.0/0.9JH & 1.53 & 0.98 & 9600/8900 \\
        2013-08-03T07:37 & 2 $\times$ 1200 & 2 $\times$ 1200 & 8 $\times$ 300 & 1.0/1.0/0.9JH & 1.55 & 0.85 & 11400/11300 \\
        2013-08-03T08:34 & 2 $\times$ 1200 & 2 $\times$ 1200 & 8 $\times$ 300 & 1.0/1.0/0.9JH & 1.49 & 0.85 & 11400/11300 \\
		
		\hline\noalign{\smallskip}
		
	\end{tabular}

\end{table*}

We determine a redshift of $z = 2.211$ from the simultaneous detection of
emission lines belonging to \lya, \oii$\lambda$3727, \hb, \oiii$\lambda$5007,
and \ha. \hb~is detected at low significance ($\sim 3 \sigma$). We show
\oiii$\lambda$5007 in the insert in Fig.~\ref{fig:line}.  \ha~is only visible in
the first epoch due to the $K$-band blocking filter used for the remaining
observations. The nebular lines exhibit a spatial extent of $\sim$ 1\farc5 and
show significant velocity structure along the slit. A drop in the continuum
bluewards of the \lya~line further supports the inferred redshift.

Using the luminosity of \ha, we can infer the star-formation rate (SFR) of the
host \citep{Kennicutt1998}. At the redshift of the GRB host, \ha~is observed
at 21\,000~\AA~where the night sky is very bright. In addition, several bright
sky-lines are superposed on the line, making an accurate estimate of the
\ha-flux difficult. We obtain a limit on the SFR by numerically integrating the
part of \ha~free of contamination and correcting for the missing flux using the
line profile and obtain $F_{H \alpha} > 4.1 \times 10^{-17} \mathrm{erg}
\mathrm{s}^{-1} \mathrm{cm}^{-2}$. After converting the \citet{Kennicutt1998}
relation to a \citet{Chabrier2003} initial mass function using
\citet{Madau2014}, we derive a limit of $SFR > 7 M_\odot$~yr$^{-1}$.
Additionally, based on the width of \oiii, an aperture covering the line is
integrated over, where synthetic sky lines \citep{Noll2012, Jones2013} have been
masked and interpolated over. From the integrated \ha-line, we estimate SFR =
$18 \pm 3$ M$_\odot$ yr$^{-1}$. From the SED-fit (Sect.~\ref{SED}), and the
detection of \lya, the host is constrained to contain very little or no dust,
although the presence of Lya does not exclude dust. Therefore we do not apply a
dust-correction to the measured \ha~flux here. \oii~is close to a region of
strong telluric absorption, which is why no SFR is inferred from this line.

The total extent of the lines in velocity space is $\sim$ 450 km/s. The line
profiles shows an asymmetric "double-horned" profile, indicating that we are
seeing a galaxy with a large degree of coherent rotational motion relative to
the line-of-sight. If we assume that we are viewing a spiral galaxy edge-on,
this is a measure of the rotational velocity of the gas. If we assume that the
spectral resolution and the turbulent width of the lines are negligible compared
to the rotational velocity, we can, based on the projected size of the source
and the width of the lines, put a constraint on the dynamical mass of the galaxy
\citep{DeBlok2014}. Based on the physical size along the slit and the velocity
width of \oiii$\lambda5007$, we infer M$_\text{dyn} \gtrsim 10^{10.8}$
M$_\odot$. Because we are viewing the host inclined at an angle relative to
edge-on and because the slit is not aligned along the long axis of the host,
this value is a lower limit. 

\subsection{Imaging observations and SED analysis} \label{SED}

In addition to the spectroscopy presented above, we imaged the field of
GRB~111117A in multiple broad-band filters using the VLT equipped with FORS2
($gRIz$ filters) and HAWK-I ($JHK_{\mathrm{s}}$ filters), long after the burst
faded. These new data are complemented by a re-analysis of some of the imaging
used in \citet{Margutti2012} and \citet{Sakamoto2013} that are available to us
(GTC $gri$-band, TNG $R$-band, and Gemini $z$-band). A log of the photometric
observations and measured brightnesses is given in
Table~\ref{tab:phot_overview}. Due to the optical darkness of this burst, see
Sect. \ref{xray}, there is likely no afterglow contamination on the measured
photometry.

All data were reduced, analyzed and fitted in a similar manner as described in
detail in \citet{Kruhler2011a} and, more recently, in \citet{Schulze2016}.
Briefly, we use our own \texttt{Python} and IRAF routines to perform a standard
reduction which includes bias/flat-field correction, de-fringing (if necessary),
sky-subtraction, and stacking of individual images. The photometry of the host
was calibrated relative to field stars from the SDSS and 2MASS catalogs in the
case of $grizJHK_{\mathrm{s}}$ filters.
For the $R$ and $I$-band photometry, we used the color transformations of
Lupton\footnote{\url{https://www.sdss3.org/dr8/algorithms/sdssUBVRITransform.php}}. 
We convert all magnitudes into the AB system, and correct for a Galactic 
foreground of $E_{B-V}=0.027~\mathrm{mag}$ \citep{Schlegel1998, Schlafly2011}.

\begin{table*}[!ht]

	\centering
	\caption{Overview of the photometric observations. \label{tab:phot_overview}}
	\begin{tabular}{ccccccc}
		\hline\hline
{Obs. Date} &  Exptime & Telescope/Instrument & Filter & Airmass & Image Quality & Host Brightness\tablefootmark{a}  \\ [1.5pt]
        \hline
{} & {ks} &    & {} & & (arcsec)  & (mag$_{\mathrm{AB}}$)  \\ [1.5pt]
		\hline
2013-08-30T07:43 & 1.45 & VLT/FORS2 & $g$ & 1.55 & 0.99 & $24.08\pm 0.09$ \\
2011-11-17T20:07 & 0.80 & GTC/OSIRIS & $g$ & 1.15 & 1.67 & $24.13\pm 0.09$ \\
2011-11-17T20:07 & 1.20 & GTC/OSIRIS & $r$ & 1.11 & 1.50 & $23.93\pm 0.08$ \\        
2013-07-17T08:37 & 1.45 & VLT/FORS2 & $R$ & 1.56 & 0.74 & $23.95\pm 0.06$ \\   
2011-11-28T21:10 & 3.60 & TNG/DOLORES & $R$ & 1.01 & 1.08 & $23.96\pm 0.13$ \\           
2011-11-17T20:07 & 0.36 & GTC/OSIRIS & $i$ & 1.08 & 1.50 & $23.89\pm 0.23$ \\   
2013-08-03T09:23 & 1.35 & VLT/FORS2 & $I$ & 1.54 & 0.93 & $24.22\pm 0.15$ \\           
2011-11-28T06:14 & 1.80 & Gemini/GMOS-N & $z$ & 1.01 & 0.84 & $24.24\pm 0.47$ \\  
2013-07-13T09:33 & 1.08 & VLT/FORS2 & $z$ & 1.49 & 0.63 & $23.76\pm 0.21$ \\             
2013-06-24T09:14 & 1.98 & VLT/HAWK-I & $J$ & 1.70 & 0.63 & $23.13\pm 0.18$ \\        
2013-06-27T09:21 & 1.68 & VLT/HAWK-I & $H$ & 1.63 & 0.91 & $22.94\pm 0.29$ \\   
2013-06-28T09:14 & 1.92 & VLT/HAWK-I & $K_\mathrm{s}$ & 1.65 & 0.76 & $23.07\pm 0.32$ \\   
\hline\noalign{\smallskip}
		
\end{tabular}

\tablefoot{
\tablefoottext{a}{All magnitudes are given in the AB system and are not corrected for the expected Galactic foreground extinction corresponding to a reddening of $E_{B-V}=0.027$\,mag.}}
\end{table*}

The multi-color SED is fit using the \citet{Bruzual2003} single stellar population
models based on a \citet{Chabrier2003} with initial mass function in
\emph{LePhare} \citep{Ilbert2006}, where the redshift is fixed to the
spectroscopic value of $z=2.211$. The best fit model is an unreddened galaxy
template, and returns physical parameters of absolute magnitude
($M_B=-22.0\pm0.1$\,mag), stellar mass ($\log(M_{\star}/M_\odot) = 9.9\pm0.2$),
stellar population age ($\tau = 0.5_{-0.3}^{+0.5}$ Gyr) and star-formation rate
($SFR_{\mathrm{SED}}=11_{-4}^{+9} M_\odot\,\mathrm{yr}^{-1}$). We show the SED
fit in Fig.~\ref{fig:SED}.

Noteworthy is the discrepancy of our new VLT/FORS2 photometry and the
re-analysis of the Gemini data to the $z$-band measurements of
\citet{Margutti2012} and \citet{Sakamoto2013}. Both of these authors report
$z$-band photometry that is brighter by 0.8~mag to 1.0 mag compared to our
value, where data taken in bluer filters are in excellent agreement. The large
$i-z$ color was mistakenly interpreted as a 4000\,\AA\,break driving the galaxy
photometric redshift of the earlier works. Using the revised photometry from
Table~\ref{tab:phot_overview}, the photometric redshift of the galaxy is
$z_{\mathrm{phot}}=2.04_{-0.21}^{+0.19}$, consistent with the spectroscopic
value at the 1~$\sigma$ confidence level.

\subsection{X-ray temporal and spectral analysis}\label{xray}

We retrieved the automated data products provided by the \textit{Swift}-XRT GRB
repository\footnote{\url{http://www.swift.ac.uk/xrt\_products/00507901}}
\citep{Evans2009}. 
The X-ray afterglow light curve can be fit with a single power-law decay with an
index $\alpha=1.27_{-0.10}^{+0.12}$. We performed a time-integrated spectral
analysis using data obtained in photon counting (PC) mode in the widest time epoch where
the $0.3-1.5\,\mathrm{keV}$ to $1.5-10\,\mathrm{keV}$ hardness ratio is constant
(namely, from $t-T_0 = 205$~s to $t-T_0 = 203.5$~ks, for a total of 29.1~ks of
data) to prevent spectral changes that can affect the X-ray column density
determination \citep{Kopac2012}. 
The obtained spectrum is well described by an absorbed power-law
model and the best-fit spectral parameters are a photon index of $2.1 \pm 0.4$ and
an intrinsic equivalent hydrogen column density $N_{\mathrm{H}}$ of $2.4_{-1.6}^{+2.4}
\times 10^{22}$~cm$^{-2}$ ($z=2.211$), assuming a solar abundance and a Galactic $N_{\mathrm{H}}$ in
the burst direction of $4.1 \times 10^{20}$~cm$^{-2}$ \citep{Willingale2013}.

A measure of the optical-to-X-ray flux ratio is parametrized in terms of the
"darkness"-parameter $\beta_{OX} $ \citep{Jakobsson2004}. Using the optical
afterglow limits \citep{Cucchiara2011, Cenko2011}, the X-ray lightcurve can be
interpolated and evaluated at the time of the non-detection. We find $\beta_{OX}
< 0.79$, consistent with what was reported in \citet{Sakamoto2013}.
\section{Reinterpretation of the restframe properties}

Because the projected distance does not change significantly between $z = 1.3$ and $z =
2.211$, all conclusions of \citet{Margutti2012} and \citet{Sakamoto2013}
relating to host offset are unaffected.

\subsection{Classification} \label{classification}

As pointed out by \citet{Margutti2012} and \citet{Sakamoto2013}, GRB~111117A is
securely classified as a sGRB. Because the observed classification indicators,
$T_{90}$ and hardness ratio, do not depend strongly on redshift
\citep{Littlejohns2013a}, the updated redshift does not change this designation.
The intrinsic spectral lag shortens, but since it is already consistent with
zero, this does not affect the classification.

\begin{figure}
	\centering
	\includegraphics[width=9cm]{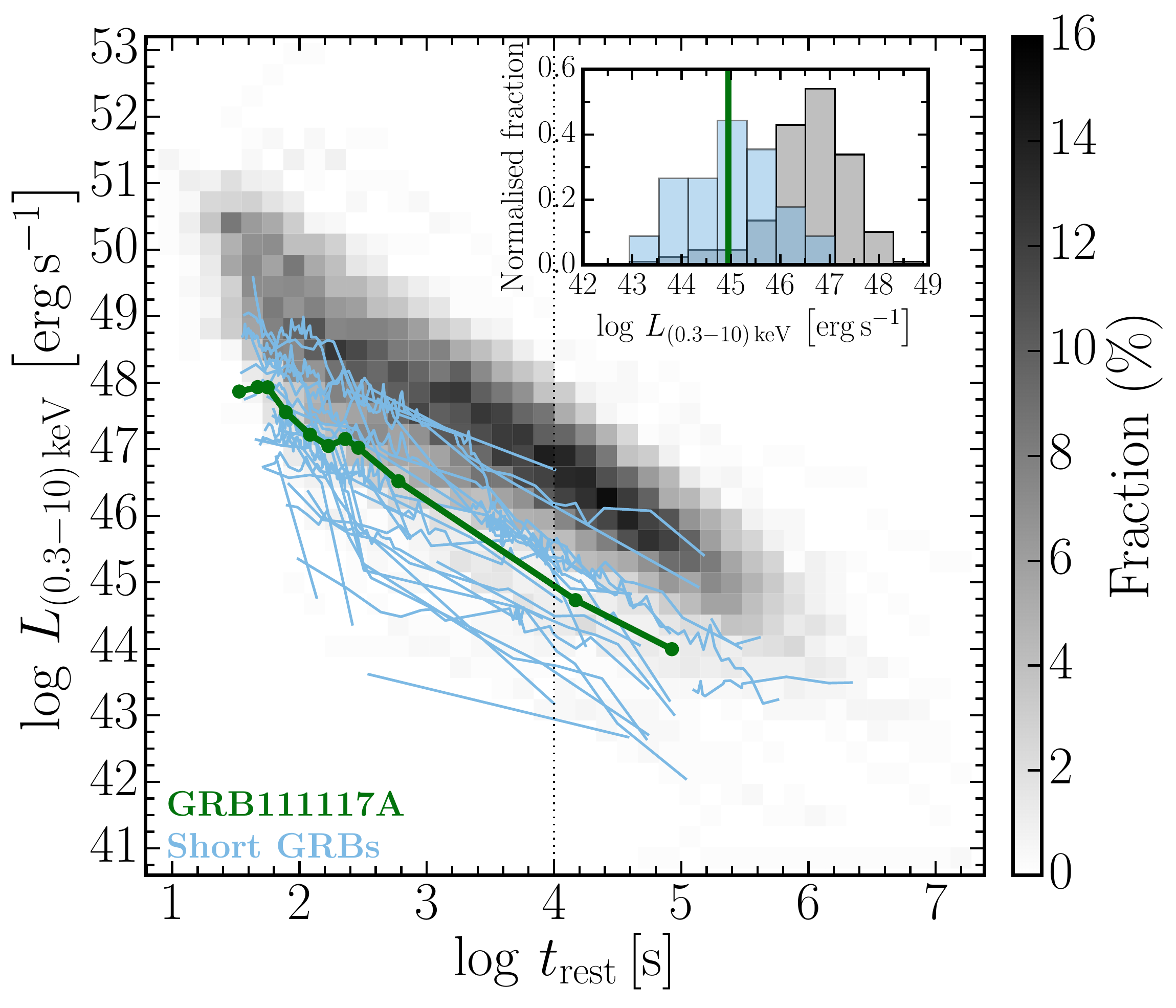}
	\caption{Restframe XRT lightcurve of GRB~111117A, compared to the general population of XRT lightcurves of GRBs. The grey shaded region is a compilation of long GRB lightcurves \citep{Evans2007, Evans2009} where the color represents density. The light blue lines are sGRB lightcurves from bursts with duration of $T_{90} \lesssim 2$ s and those that were classified as short in \citet{Kann2011, Berger2014, DAvanzo2014a}. The thick green line is GRB~111117A. Despite the remarkably high redshift, the luminosity is comparable to the bulk of the short burst population and subluminous compared to the lGRB population.}
	\label{fig:sxray_lightcurve}
\end{figure}

The intrinsic luminosity is shown in the X-ray light curve (Fig.
\ref{fig:sxray_lightcurve}) and it is sub-luminous compared to the majority of
long GRBs. The inset in Fig. \ref{fig:sxray_lightcurve} shows the luminosity
distribution at 10 ks. The sub-samples comprise of 333 long, 19 short GRBs, and
GRB~111117A. The mean and the 1-$\sigma$ dispersions of the samples are
L$_{lGRB} = 46.59 \pm 0.87$ and L$_{sGRB} = 44.96 \pm 0.94$. GRB~111117A had a
luminosity of 44.95 at 10 ks which is very close to the peak of the sGRB
luminosity distribution at 10 ks, but an outlier from the lGRG distribution,
further supporting the short classification.

\citet{Bromberg2013} investigated the degree to which the long and short
population distributions overlap and quantified the certainty in class
membership. According to \citet{Bromberg2013}, GRB~111117A has $96_{-5}^{+3}$
percent probability of being a sGRB. Compared to the other two sGRB candidates
at high redshift, GRB~060121 \citep{DeUgartePostigo2006, Levan2006} at $1.7
\lesssim z \lesssim 4.5$ ($17_{-15}^{+14}$ per cent) and GRB~090426
\citep{Antonelli2009, Levesque2010, Thone2011} at $z = 2.609$ ($10_{-10}^{+15}$
per cent), the certainty in class membership for GRB~111117A is much higher.

Additionally, \citet{Horvath2010} classifies both GRB~060121 and GRB~090426 as
intermediate duration bursts. This comes from both events having very soft
spectra, as compared to the hard ones typically seen in sGRBs. Intermediate
bursts are very clearly related in their properties to lGRBs
\citep{DeUgartePostigo2011}, so they are unlikely to come from compact object
mergers.

The redshift and secure classification of GRB~111117A mean that it occurred
when the universe was younger by 1.8 Gyr compared to any other non-collapsar GRB
ever detected. This number is 3.2 Gyr for the next-highest spectroscopic
redshift. If the merger of NSs is the primary agent for the $r$-process element
enrichment of the universe \citep{Goriely2011, Ji2016, Komiya2016}, this
marks the earliest detection of this process.

\subsection{Restframe $N_\mathrm{H}$} \label{restnH}

\begin{figure}
	\centering
	\includegraphics[width=9cm]{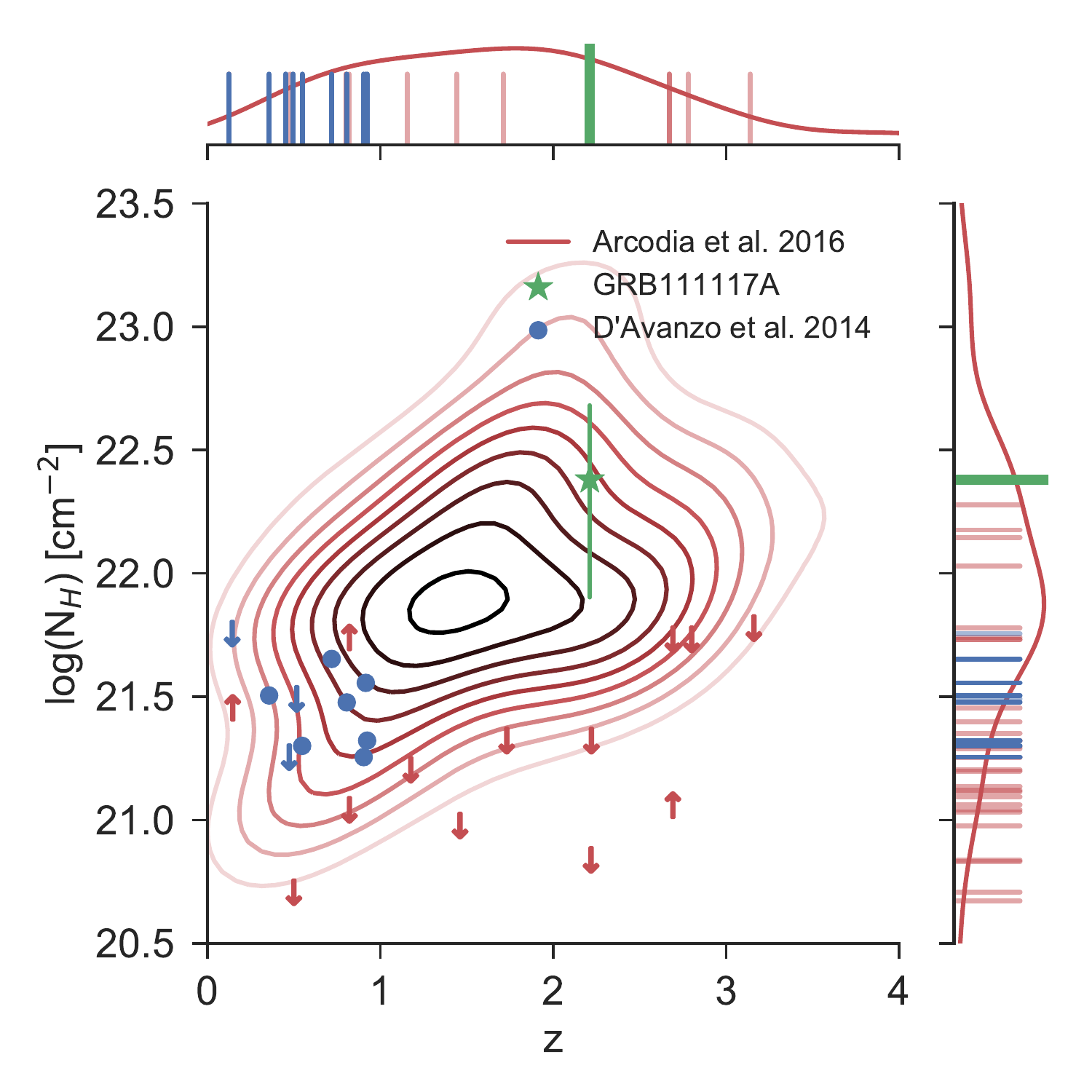}
	\caption{Rest frame, X-ray derived equivalent hydrogen column densities for GRB~111117A compared to complete samples of both long and short populations of GRBs. The detections are replaced with contours and the limits are shown with arrows. Marginalizations over both axes are shown where the limits are shown as transparent bars and detections as solid ones.}
	\label{fig:NH_z}
\end{figure}

We show the recalculated $N_\mathrm{H}$ in Fig.~\ref{fig:NH_z} where we compare
with the distributions of complete samples of both long and short GRBs. The lGRB
sample is from \citet{Arcodia2016} and the sGRB sample is from
\citet{DAvanzo2014a}. From the sGRB sample of \citet{DAvanzo2014a} we have
excluded GRB~090426 which does likely not belong in a short sample as highlighted in
Sect. \ref{classification}. Both comparison samples consider the largest
temporal interval of constant hardness ratio to prevent spectral changes that
can affect the X-ray derived column density. The 17 (5) of the 99 (15) long
(short) GRBs which do not have redshift have been excluded from our analysis.

GRB~111117A occupies a unique position in Fig.~\ref{fig:NH_z} with the highest
$N_H$ and highest redshift of all sGRBs. The short sample, excluding
GRB~111117A, is located at low redshifts ($z < 1$) and is found to populate a
column density environment similar to that of lGRBs at comparable redshifts
\citep{DAvanzo2014a}. The inferred hydrogen column density for GRB~111117A seems
to follow the trend with increasing $N_H$ as a function of redshift as found for
the lGRB afterglows \citep{Campana2010, Starling2013, Arcodia2016}. This is
related to what is found by \citet{Kopac2012, Margutti2013}, that $N_H$ seems to
be comparable for long and short GRBs when compared at similar redshifts.

The redshift evolution of $N_H$ in the hosts of lGRBs is not reproduced by
\citet{Buchner2017}, using a different $N_H$ inference methodology. Instead a
correlation between $N_H$ and host stellar mass is suggested. Assuming that the
different $N_H$-fitting methodologies yield comparable results, GRB~111117A is
an outlier from the relation suggested by \citet{Buchner2017} by more than the
intrinsic scatter, although some lGRB hosts populate a similar region in the
$N_H$-$M_\star$ relation. Additionally, for lGRBs, $N_H$ correlates with the
surface luminosity at the explosion site \citep{Lyman2017}. The large offset of
GRB~111117A relative to the host center derived in \citet{Margutti2012} and
\citet{Sakamoto2013} is difficult to reconcile with galaxy-scale gas providing
the X-ray absorption. Along with the absence of dust, the large offset from the
host center indicates that the high $N_H$ arises because the density in the GRB
surrounding is high, or because the light from the afterglow traverses a region
of dense gas. The optical darkness of the GRB is an additional hint of the high
density in the GRB surroundings or along the line of sight.

\subsection{Host galaxy}

From the clear host association, GRB~111117A does not belong to the hostless
class of GRBs \citep{Berger2010a} and because the host exhibits emission lines
this is indicative of a population of relatively young stars. As the majority of
sGRBs \citep{Fong2013b}, the host of GRB~111117A is therefore a late-type galaxy
and is entirely consistent in terms of stellar mass and stellar age with the
general population of sGRB hosts ($\left\langle M _* \right\rangle = 10^{10.1}
M_{\odot}$ and $\left\langle \tau _* \right\rangle = 0.3 $~Gyr)
\citep{Leibler2010}. Being a late-type host, both the stellar mass and sSFR are
entirely within the range expected for the hosts of sGRBs \citep{Behroozi2014}.
Our constraints on the dynamical mass is also well accommodated by the expected sGRB
host halo mass \citep{Behroozi2014}.

The SFR is $\sim$1 order of magnitude higher than the typical SFR for sGRB host
galaxies \citep{Berger2014} and more similar to the SFR found in the hosts of
lGRBs at a corresponding redshift \citep{Kruhler2015}. Only two hosts in the
sample of short GRBs compiled in \citet{Berger2014} have a more vigorous star
formation, meaning that it is in the very upper end of the star formation
distribution. The cosmic SFR evolution of the universe likely plays a role due
to the proximity of GRB~111117A to the peak of cosmic SFR \citep{Madau2014}.
The high SFR is partly a selection effect, because a less star forming galaxy
would exhibit weaker emission lines, thus making the redshift harder to
determine. Additionally, it is natural to expect some evolution in the hosts of
sGRBs with redshift as illustrated for $N_H$ in Sect.~\ref{restnH}.

The detection of \lya~is consistent with the SED-inferred absence of dust,
despite the moderate stellar age which would suggest the opposite. The centroid
of the Ly$\alpha$ emission is found to be redshifted by $\sim$~220~km~s$^{-1}$
with respect to systemic, which is similar to what is found for long GRB hosts
\citep{Milvang-Jensen2012a} where the outflow is attributed to star formation.


\section{Implications for the redshift distribution of sGRBs}

A single sGRB at high redshift does little in terms of constraining the redshift
distribution of sGRBs. In particular, other sGRB hosts could be missed because
they are intrinsically fainter and thus the high redshift of GRB~111117A is only
measured due to the brightness of its host. \citet{Berger2014} compiled a
sample of sGRB host luminosities, normalized by the characteristic galaxy
luminosity at their respective redshift, $L_B/L^{\star}_{B}$. To convert the
SED-inferred $M_B$ of GRB~111117A to $L_B/L^{\star}_{B}$, we use the
characteristic absolute $B$-band magnitude of the Schechter function for blue
galaxies ($U - V < 0.25$) in the redshift window $2.0 \leq z \leq 2.5$ from
\citet{Marchesini2007} and obtain $L_B/L^{\star}_{B} = 1.2$.

Using the complete, flux-limited selection of bursts from \citet{DAvanzo2014a},
excluding GRB~111117A and the likely non-sGRB GRB~090426, we have a
statistically homogeneous sample from which we can address the implications of
the redshift of GRB~111117A. Out of the 14 hosts in the sample, 10 (71 per cent)
have measured redshifts and $L_B/L^{\star}_{B}$. Of the complete sample, the
host of GRB~111117A  is brighter than 80 per cent of the hosts with measured
$L_B/L^{\star}_{B}$. Even if we conservatively assume that \textit{all} the
hosts missing $L_B/L^{\star}_{B}$ are brighter than the host of GRB~111117A, the
host is still brighter than > 60 per cent of sGRB hosts. For all 26 hosts with
$L_B/L^{\star}_{B}$ from \citet{Berger2014}, the host of GRB~111117A is brighter
than 73 per cent.

If we assume that we are able to obtain emission-line redshifts from hosts, 0.5
mags fainter \citep[$R < 24.5$~mag;][]{Kruhler2012}, then we would have missed
60 per cent of the redshifts (6 out of 10 hosts), due too the host being too
faint, were they at the redshift of GRB~11117A. The corresponding number is
around 45 per cent (12 out of 26) from the full sample of \citet{Berger2014},
reflecting the lower mean $L_B/L^{\star}_{B}$ of the complete sample. Because
the average SFR of galaxies hosting lGRBs is higher than for galaxies hosting
sGRBs, the fraction of missed burst redshifts is likely higher although the
cosmic SFR evolution could play a role in improving redshift determinability.

A fraction of the bursts missing redshift are host-less but appear to be
spatially correlated with galaxies that are likely at moderate redshifts
\citep{Tunnicliffe2014}, but should some of the remainder be at high redshift,
the missed fraction will increase. If we assume that \textit{all} the bursts
that are missing redshifts are at high-$z$ and missed due to host faintness, 10
out of 14 hosts in the complete sample (71 per cent) would be missed at $z =
2.211$. This serves as an upper limit on the fraction of missed burst redshifts
at high-$z$. Conversely, if all bursts missing redshift are at low redshift and
missed for other reasons, 6 out of 14 hosts (43 per cent) would be missed at $z
= 2.211$. The two limits indicate that we would miss between 43 and 71 per cent
of sGRB hosts at $z \sim 2$ due to host faintness.

The theoretical redshift distribution of sGRBs depends on the type of delay-time
function used to model the progenitor system. The likelihood preferred lognormal
time delay models investigated by \citet{Wanderman2015} predict a sGRB rate at
$z = 2.211$, $\sim$ two orders of magnitude lower compared to the peak rate at
$z = 0.9$. According to \citet{Wanderman2015}, this preference depends
critically on the absence of non-collapsar sGRBs at $z \gtrsim 1.2$. The higher
determined redshift of GRB~111117A, and the likely number of additional high-z
sGRB could change the preferred time delay models. The redshift of GRB~111117A,
on the other hand, is close to the expected peak in sGRB rate calculated using
the power law delay time models \citep{Behroozi2014, Wanderman2015,
	Ghirlanda2016}, meaning we could be missing a large fraction of sGRBs.

\section{Constraints on progenitor separation}

At $z = 2.211$, the age of the universe is 3 Gyr. If the progenitor systems of
sGRBs are the merger of two NSs, this sets a hard upper limit to the coalescence
timescale for such a system. In the absence of other mechanisms, the timescale
of the orbital decay of the system is set by the energy loss due to
gravitational waves, which in turn is set by the mass of the constituent compact
objects, the eccentricity of the orbit and the separation of the two \citep{Postnov2014}. If we assume that the
formation timescale of the first galaxies is short compared to the time since
the Big Bang \citep{Richard2011}, and if we assume a mass of 1.4~$M_\odot$ for
each of the NSs in a circular orbit at the time of system formation, this places
a hard upper limit on the initial separation, of $a_0 < 3.2~R_\odot$.

In practice most NS-NS binaries will be eccentric at formation because of the SN
natal kicks. For more eccentric orbits, the coalescence timescale decreases,
leading to a decrease in the constraint on the initial separation and larger
initial separation constraints. As noted by \citet{Postnov2014}, it takes
eccentricities > 0.6 to significantly shorten the merger time.

Using the inferred stellar population age from our SED fit, then we obtain a
(softer) limit on the initial separation of $a_0 < 2.1~R_\odot$. However, this
does not account for the possibility there could be an underlying stellar
population of older stars from a previous star-formation epoch. To investigate
the possible impact of the presence of an old stellar population, we followed
\citet{Papovich2001} and re-fitted the observed SED with the best-fit template
to which an additional stellar population of old stars was added. For each
galaxy, this old population was set as the SPSs with same parameters that the
best-fit SED excepts the age, which was set to the age of the Universe at the
observed redshift. In principle, this can constrain the maximum contribution of
old populations within the photometric error bars \citep[see][ for
details]{Papovich2001}. We find a negligible contribution to the stellar mass
(i.e. variations much smaller than the statistical uncertainty associated with
the best-fit template). The delay time between formation and explosion is well
accommodated by the models of \citet{Belczynski2006}, although the longest delay
times are excluded. This is especially true given the late type nature of the
host \citep{OShaughnessy2008}.

\section{Conclusions}

We have here provided a revised, spectroscopic redshift measurement for the
short GRB~111117A based on host galaxy emission lines setting it at $z = 2.211$.
This value supersedes the previous photometric redshift of \citet{Margutti2012}
and \citet{Sakamoto2013}. The erroneous redshift estimate of previous authors is
attributed to a discrepancy in the measured $z$-band magnitude, and highlights
the importance of deep spectroscopic studies of sGRB hosts at medium resolution.

Using the new distance, the X-ray derived $N_H$ towards GRB~111117A is the
highest within a complete sample of sGRB hosts and is consistent with the
$N_H-z$ evolution traced by the hosts of lGRBs. The SFR of the host is in the
upper end of the sGRB host SFR distribution and no significant amount of dust is
present. The high $N_H$ is difficult to reconcile with the large projected host
offset and the absence of dust. One possible explanation could be, that
GRB~111117A is formed through a prompt channel of sGRB formation and originates
in a star forming region located in the outskirts of the host.

Although a single burst carries little leverage in terms of constraining the
redshift distribution of sGRB, the high redshift of GRB~111117A needs to be
accommodated in progenitor models. A lognormal delay time model predicts a very
low volumetric density of bursts at $z = 2.211$, whereas a power law delay time
model peaks near GRB~111117A. If more sGRBs are at similarly high redshifts, but
are missed due to the faintness of their hosts, a lognormal delay time model
will be disfavored. Compared to a sample of sGRB hosts, GRB~111117A is more
luminous than 80 per cent of a complete sample of sGRB hosts with measured
luminosities. Assuming a host brightness redshift determination threshold, for
between 43 and 71 per cent of sGRB hosts, we would be unable to determine a
redshift should they be at a similar redshift as GRB~111117A. This could
indicate that potentially, a significant fraction of sGRBs are at high $z$ but
missed due to host faintness.

Using the age of the universe at the time of explosion allows us to set
constraints on the maximal separation between the engine constituents at the
time of formation. We find that the maximal separation of two NSs at system
formation time is $a_0 < 3.2~R_\odot$, which excludes some of the formation
channels with the longest timescales.

All data, code and calculations related to the paper along with the
paper itself are available at \url{https://github.com/jselsing/GRB111117A}.

\begin{acknowledgements}
We thank Jens Hjorth and Lise Christensen for useful discussions regarding the interpretation of this event. We thank Mathieu Puech for testing the possible contribution from an older stellar population in the SED.
TK acknowledges support through the Sofja Kovalevskaja Award to P. Schady.
SDV is supported by the French National Research Agency (ANR) under contract ANR-16-CE31-0003 BEaPro 
JJ acknowledges support from NOVA and NWO-FAPESP grant for advanced
instrumentation in astronomy.
NRT and KW acknowledge support from STFC Consolidated
Grant ST/N000757/1
CT acknowledges support from a Spanish National Research Grant of Excellence
under project AYA 2014-58381-P and funding associated to a Ramón y Cajál
fellowship under grant number RyC-2012-09984.
AdUP acknowledges support from a Ramón y Cajal fellowship, a BBVA Foundation
Grant for Researchers and Cultural Creators,and the Spanish Ministry of Economy
and Competitiveness through project AYA2014-58381-P. Partly based on
observations made with the Gran Telescopio Canarias (GTC).
ZC acknowledges support from the Spanish research project AYA 2014-58381-P and
support from Juan de la Cierva Incorporaci\'on fellowships IJCI-2014-21669.
RSR acknowledges AdUP's BBVA Foundation Grant for Researchers and Cultural
Creators and support from ASI (Italian Space Agency) through the Contract n. 2015-046-R.0 and from European Union Horizon 2020 Programme under the AHEAD project (grant agreement n. 654215).
This research made use of Astropy, a community-developed core Python package for Astronomy \citep{TheAstropyCollaboration2013}. The analysis and plotting has been achieved using the Python-based packages Matplotlib \citep{Hunter2007}, Numpy and Scipy \citep{VanderWalt2011} along with other community-developed packages.
This work made use of observations obtained with the Italian 3.6 m Telescopio Nazionale Galileo (TNG) operated on the island of La Palma by the Fundaci\'on Galileo Galilei of the INAF (Istituto Nazionale di Astrofisica) at the Spanish Observatorio del Roque de los Muchachos of the Instituto de Astrof\'isica de Canarias.
Based on data from the GTC Archive at CAB (INTA-CSIC) and on observations obtained at the Gemini Observatory, which is operated by the Association of Universities for Research in Astronomy, Inc., under a cooperative agreement with the NSF on behalf of the Gemini partnership: the National Science Foundation (United States), the National Research Council (Canada), CONICYT (Chile), Ministerio de Ciencia, Tecnología e Innovación Productiva (Argentina), and Ministério da Ciência, Tecnologia e Inovação (Brazil).

\end{acknowledgements}

\bibliographystyle{aa_arxiv}
\bibliography{GRB111117A}

\end{document}